\documentclass[a4paper,11pt]{article}
\usepackage{a4wide}
\usepackage{amsmath}
\usepackage{amssymb}
\usepackage[caption=false,font=footnotesize]{subfig}

\usepackage[all,graph,curve]{xy}

\newtheorem{thm}{Theorem}
\newtheorem{lem}[thm]{Lemma}

\newenvironment{pf}[1][Proof]{\begin{trivlist}
\item[\hskip \labelsep {\bfseries #1}]}{\end{trivlist}}
\newenvironment{definition}[1][Definition]{\begin{trivlist}
\item[\hskip \labelsep {\bfseries #1}]}{\end{trivlist}}

\newcommand{\qed}{\nobreak \ifvmode \relax \else
      \ifdim\lastskip<1.5em \hskip-\lastskip
      \hskip1.5em plus0em minus0.5em \fi \nobreak
      \vrule height0.75em width0.5em depth0.25em\fi}

\begin{document}

\newcommand{\name}[1]{\textsc{#1}}

\title{Determinant Sums for Undirected Hamiltonicity}

\author{Andreas Bj\"orklund}

\date{}


\maketitle 

\begin{abstract}
We present a Monte Carlo algorithm for Hamiltonicity detection in an $n$-vertex undirected graph running in $O^*(1.657^{n})$ time.
To the best of our knowledge, this is the first superpolynomial improvement on the worst case runtime for the problem since the $O^*(2^n)$ 
bound established for TSP almost fifty years ago (Bellman 1962, Held and Karp 1962).
It answers in part the first open problem in Woeginger's 2003 survey on exact algorithms for NP-hard problems.

For bipartite graphs, we improve the bound to $O^*(1.414^{n})$ time. Both the bipartite and the general algorithm can be implemented to use space polynomial in $n$. 

We combine several recently resurrected ideas to get the results.
Our main technical contribution is a new reduction inspired by the 
algebraic sieving method for $k$-Path (Koutis ICALP 2008, Williams IPL 2009). 
We introduce the Labeled Cycle Cover Sum in which we are set to count weighted arc labeled cycle covers over a finite field of characteristic two. 
We reduce Hamiltonicity to Labeled Cycle Cover Sum and apply the determinant summation technique for Exact Set Covers (Bj\"orklund STACS 2010) to evaluate it.

\end{abstract}


\section{Introduction}
An undirected graph $G=(V,E)$ on $n$ vertices is said to be Hamiltonian if it has a Hamiltonian cycle, a vertex order $(v_0,v_1,\cdots, v_{n-1})$ such that $v_iv_{i+1}\in E$ for all $i$. The indices are enumerated modulo $n$ requiring also that $v_{n-1}v_0$ is an edge.
The problem of detecting if a graph is Hamiltonian is called the \name{Hamiltonicity} problem and was one of the first identified as NP-hard. It is on Karp's original list \cite{K72}, but is perhaps best known as a special case of the Traveling Salesman Problem (\name{TSP}). The \name{TSP} asks for a tour visiting every vertex of an edge weighted graph exactly once that minimizes the total weight.
Bellman~\cite{B60,B62} and independently Held and Karp~\cite{HK62}  described in the early 1960's a dynamic programming recurrence that solves the general \name{TSP} in $O(n^22^n)$ time. Their bound also holds for the special case of \name{Hamiltonicity} and is still the strongest known. 
Under the widely acknowledged Exponential Time Hypothesis, the \name{Hamiltonicity} problem has $exp(\Omega(n))$ runtime \cite{IPZ01}.
There is however no known reason to expect the exponential base to be precisely two.
Woeginger in his survey on exact algorithms for NP-Hard problems ~\cite{W03} observes this and asks in Open problem 3.1 for a $O^*(c^n)$ time algorithm for \name{TSP} and \name{Hamiltonicity} for some $c<2$. $O^*(f(n))$ suppresses polylogarithmic functions in $f(n)$. We solve the latter problem. 

\begin{thm}
\label{thm: gen}
There is a Monte Carlo algorithm detecting whether an undirected graph on $n$ vertices is Hamiltonian or not running in $O^*(1.657^n)$ time, with no false positives and false negatives with probability exponentially small in $n$.
\end{thm}

For graphs having an induced subgraph with many disconnected components, most notably bipartite graphs, we get an even stronger bound.

\begin{thm}
\label{thm: bip}
There is a Monte Carlo algorithm detecting whether an undirected graph on $n$ vertices with a given independent set of size $i$ is Hamiltonian or not running in $O^*(2^{n-i})$ time, with no false positives and false negatives with probability exponentially small in $n$.
\end{thm}

We also note that our algorithm can be used to solve \name{TSP} with integer weights via self-reducibility at the cost of a runtime blow-up by roughly 
a factor of the sum of all edges' weights.

\begin{thm}
\label{thm: TSP}
There is a Monte Carlo algorithm finding the weight of the lightest TSP tour in a positive integer edge weighted graph on $n$ vertices in $O^*(w1.657^n)$ time, where $w$ is the sum of all weights, with error probability exponentially small in $n$.
\end{thm}

\subsection{Previous Work}
Bellman's~\cite{B60,B62} and Held and Karp's~\cite{HK62} algorithm for \name{TSP} in an $n$-vertex complete graph $G=(V,E)$ with edge weights $\ell:E\rightarrow \mathbb{R}^+$ is based on defining
$w_{s,t}(X)$ for $s,t\in X\subseteq V$ as the weight of the lightest path from $s$ to $t$ in the induced graph $G[X]$ visiting all vertices in $X$ exactly once. This quantity obeys the simple recursion
\begin{displaymath}
w_{s,t}(X)\!=\!
\left\{\!\!\! \begin{array}{ll} \min_{u\in X\setminus \{s,t\}} w_{s,u}(X\setminus\{t\})+\ell(ut) &\!\!\!\!:|X|>2 \\
\ell(st)&\!\!\!\!:|X|=2 \\
\end{array} \right.
\end{displaymath}

Using bottom-up dynamic programming with $s$ fixed, the lightest tour can be evaluated by $\min_{t\in V\setminus\{s\}} w_{s,t}(V)+\ell(st)$ in total $O(n^22^n)$ time.
An \name{Hamiltonicity} instance $G$ can naturally be embedded in a \name{TSP} instance on the same number of vertices. 
Simply let the weight function $\ell$ take the value $0$ for vertex pairs corresponding to an edge in $G$, and $1$ otherwise.

Another algorithm amenable to \name{Hamiltonicity} with (almost) the same running time is the inclusion--exclusion counting over $n$-long closed walks in the induced subgraphs. The algorithm has been (re)discovered at least three times \cite{KGK77,K82,B93}.
Let $s$ be any vertex in the graph, then the number of Hamiltonian cycles is given by
\begin{displaymath}
\sum_{X\subseteq V\setminus\{s\}} (-1)^{|V\setminus(X \cup \{s\})|}(\mathbf{A}[X\cup\{s\}]^n)_{s,s}
\end{displaymath}
Here, $\mathbf{A}[Y]$ denotes the adjacency matrix of the induced graph $G[Y]$, and $(\mathbf{A}[Y]^n)_{s,s}$ the entry at row and column $s$ of the matrix 
$\mathbf{A}[Y]^n$.
The idea behind the algorithm is that crossing walks will be canceled since they are counted equally many times with the sign factor $+1$ as with $-1$.

In restricted graph classes the general $O(2^n)$ bound has been sharpened. Broersma et al.~\cite{BFPP09} proved that \name{Hamiltonicity} in claw--free graphs has an $O^*(1.682^n)$ time algorithm. Iwama and Nakashima~\cite{IN07} improving slightly on Eppstein~\cite{E07}, showed that TSP in cubic graphs admits an $O^*(1.251^n)$ time algorithm. In graphs of maximum degree four, Gebauer~\cite{G08} described how to count the Hamiltonian cycles in $O^*(1.715^n)$ time. For larger degrees, only minuscule improvements are known.
Bj\"orklund et al.~\cite{BHKK08} observed that both the Bellman-Held-Karp dynamic programming and the inclusion--exclusion algorithm need to look only at $X$ for which $G[X]$ is connected and include $s$. These are at most $(2-\epsilon)^n$ with $\epsilon$ depending inversely exponentially on the maximum degree.

Another line of research addresses a natural parameterized version of the problem called the $k$-\name{Path} problem: how much time is required to find a simple (noncrossing) path on $k$ vertices in an $n$ vertex graph.
Alon et al. \cite{AYZ95} showed the first $c^kn^{O(1)}$ time algorithm for some constant $c$ for the problem. The constant $c$ has since been improved several times culminating in the work of Koutis~\cite{K08} introducing an interesting algebraic sieving technique. His algorithm was
subsequently refined by Williams~\cite{W09} to yield a $2^kn^{O(1)}$ time algorithm for the $k$-\name{Path} problem. In the extreme $k=n$, their algorithm's runtime coincides with
the previously best $O^*(2^n)$ time bound for \name{Hamiltonicity}.

\subsection{Our Approach}

The inclusion--exclusion algorithm in the previous section has several desirable properties:
It uses space polynomial in the input size, it works also for directed graphs, it is deterministic, and it is capable of counting the solutions.
Our algorithm also manages with only polynomial space after some extra care, but does not have the other three properties. Indeed,
we crucially depend on the graph being undirected and that polynomial identity testing has an efficient randomized algorithm (whereas no deterministic is known). Moreover, our strategy does not even seem to be able to approximate the number of solutions.

The inclusion--exclusion algorithm can be thought of as first counting too much (all closed $n$-walks through $s$) and then canceling out every false contribution (crossing $n$-walks). We will take a similar approach, but count instead weighted cycle covers in directed graphs over fields of characteristic two. A cycle cover in a directed $n$-vertex graph is a set of $n$ arcs such that every vertex is the origin of one arc, and the end of another. The arcs together describe disjoint cycles covering all vertices of the graph. In particular, the cycle covers contain the Hamiltonian cycles.

Our algorithm is much inspired by the recent work of Koutis~\cite{K08} and Williams~\cite{W09} for $k$-\name{Path}. Although we don't apply their work directly, we use several of their ideas. In particular we evaluate multivariate polynomials over fields of characteristic two to sieve for the Hamiltonian cycles, just as they do for $k$-paths.

The main new technical ingredient is the introduction of determinants to count weighted cycle covers. This is an extension of the idea of using determinants to count perfect matchings recently employed by Bj\"orklund \cite{B10} for \name{Exact Set Cover}. Unlike Koutis--Williams we are unable to construct small arithmetic circuits. Instead we depend on the efficient algorithms for computing a matrix determinant numerically.

\subsection{Organization}
The rest of the paper is organized as follows. In Section~\ref{sec: prel} we introduce the technical machinery needed. In particular we define our weighted cycle cover problem, hint at  how it relates to the \name{Hamiltonicity} problem, and presents a way to compute it.
In Section~\ref{sec: red} we describe how we can reduce a \name{Hamiltonicity} instance to the weighted cycle cover problem and prove our main claims 
Theorem~\ref{thm: gen} and ~\ref{thm: bip} given exponential space. 
In Section~\ref{sec: pspace} we argue how the algorithm can be modified to use only polynomial space. Finally, in Section~\ref{sec: TSP} we note how the technique can be extended in a known way to solve TSP to prove Theorem~\ref{thm: TSP}.

\section{Preliminaries}
\label{sec: prel}
In a directed graph $D=(V,A)$ a cycle cover is a subset $C\subseteq A$ such that for every vertex $v\in V$ there is exactly one arc $a_{v1}\in C$ starting in $v$, and exactly one arc $a_{v2}\in C$ ending in $v$. The graphs in this paper have no loops, i.e. arcs connecting a vertex to itself, and thus we also have that $a_{v1}\neq a_{v2}$. We denote by $cc(D)$ the family of all cycle covers of $D$, and by
$hc(D)\subseteq cc(D)$ the set of Hamiltonian cycle covers. A Hamiltonian cycle cover consists of one big cycle passing through all vertices. 
The remaining cycle covers (which have more than one cycle), $cc(D)\setminus hc(D)$, are called non-Hamiltonian cycle covers. 
Although an element of $hc(D)$ is formally a subset of arcs, we will sometimes write it as a vertex order $(v_0,v_1,\cdots,v_{n-1})$ 
implicitly referring to the arcs $v_iv_{i+1}$ as the actual Hamiltonian cycle. 
For undirected graphs $G$, $hc(G)$ includes the Hamiltonian cycles with orientation, i.e. traversed in both directions. 
Hence we will for a Hamiltonian cycle $H\in hc(G)$ for an undirected graph $G$, talk about \emph{arcs} $uv\in H$ inferring that the cycle is oriented from $u$ to $v$ along the edge $uv$.

We write $g:A\twoheadrightarrow B$ for a surjective function $g$ from the domain $A$ to the codomain $B$.
For a function $g:A\twoheadrightarrow B$ we associate the function $g^{-1}:B\rightarrow 2^A$ as its preimage, $g^{-1}(b)=\{\,a\in A\colon g(a)=b\}$. For a matrix $\mathbf{A}$, we denote by $\mathbf{A}_{i,j}$ the element at row $i$ and column $j$. For a polynomial $p(r)$ in an indeterminate $r$ we write $[r^l]p(r)$ to address the coefficient of the monomial $r^l$ in $p(r)$.

We will reduce \name{Hamiltonicity} to a variant of cycle cover counting defined next. We introduce the \name{Labeled Cycle Cover Sum}. The name stems from the fact that every arc in the cycle cover is labeled by a nonempty subset of a set of labels. 

\begin{definition}
The \name{Labeled Cycle Cover Sum}
for a directed graph $D=(V,A)$, a label set $L$, and a function $f:A\times 2^{L}\setminus\{\emptyset\} \rightarrow R$ on some codomain ring $R$ is
\begin{equation}
\label{def: lcc}
\Lambda(D,L,f) = \sum_{C\in cc(D)} \sum_{g:L\twoheadrightarrow C} \prod_{a\in C} f(a,g^{-1}(a)).
\end{equation}
\end{definition}

Note in particular that the inner sum is over all surjective functions $g$, meaning that the label $g^{-1}(a)$ is a nonempty subset of $L$ for all arcs $a\in C$.
In words the computation is over all arc labeled cycle covers of the graph such that all arc labels are nonempty, are pairwise disjoint, and together exhaust all of the labels $L$.

\subsection{Cycle Cover Cancelation in Characteristic Two}
\begin{figure}
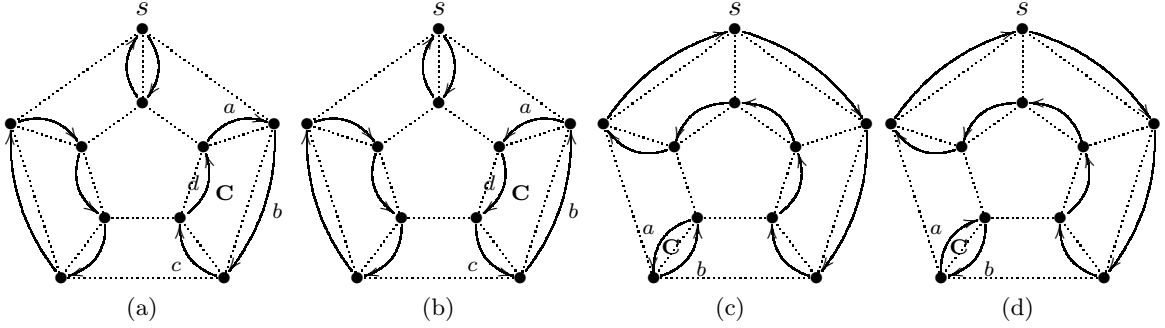

\centering
\subfloat[]{
\xygraph{
!{<0cm,0cm>;<0cm,1.4cm>:<1.4cm,0cm>::}
!{(0,0);a(0)**{}?(0.6)}*{\bullet}="f"
!{(0,0);a(72)**{}?(0.6)}*{\bullet}="g"
!{(0,0);a(144)**{}?(0.6)}*{\bullet}="h"
!{(0,0);a(216)**{}?(0.6)}*{\bullet}="i"
!{(0,0);a(288)**{}?(0.6)}*{\bullet}="j"
!{(0,0);a(0)**{}?(1.5)}*{s}
!{(0,0);a(0)**{}?(1.3)}*{\bullet}="a"
!{(0,0);a(72)**{}?(1.3)}*{\bullet}="b"
!{(0,0);a(144)**{}?(1.3)}*{\bullet}="c"
!{(0,0);a(216)**{}?(1.3)}*{\bullet}="d"
!{(0,0);a(288)**{}?(1.3)}*{\bullet}="e"
"a"-@{.}"b" "b"-@{.}"c" "c"-@{.}"d" "d"-@{.}"e" "e"-@{.}"a"
"a"-@{.}"f" "b"-@{.}"g" "c"-@{.}"h" "d"-@{.}"i" "e"-@{.}"j"
"j"-@{.}"i" "i"-@{.}"h" "h"-@{.}"g" "g"-@{.}"f" "f"-@{.}"j"
"a":@/^/"f" "f":@/^/"a"
"e":@/^/"j" "j":@/_/"i" "i":@/^/"d" "d":@/^/"e"
"g":@/^/"b" ^{a} "b":@/^/"c" ^{b} "c":@/^/"h" ^{c} "h":@/_/"g" ^{d} _{\mathbf{C}}
}
}
\subfloat[]{
\xygraph{
!{<0cm,0cm>;<0cm,1.4cm>:<1.4cm,0cm>::}
!{(0,0);a(0)**{}?(0.6)}*{\bullet}="f"
!{(0,0);a(72)**{}?(0.6)}*{\bullet}="g"
!{(0,0);a(144)**{}?(0.6)}*{\bullet}="h"
!{(0,0);a(216)**{}?(0.6)}*{\bullet}="i"
!{(0,0);a(288)**{}?(0.6)}*{\bullet}="j"
!{(0,0);a(0)**{}?(1.5)}*{s}
!{(0,0);a(0)**{}?(1.3)}*{\bullet}="a"
!{(0,0);a(72)**{}?(1.3)}*{\bullet}="b"
!{(0,0);a(144)**{}?(1.3)}*{\bullet}="c"
!{(0,0);a(216)**{}?(1.3)}*{\bullet}="d"
!{(0,0);a(288)**{}?(1.3)}*{\bullet}="e""a"-@{.}"b" "b"-@{.}"c" "c"-@{.}"d" "d"-@{.}"e" "e"-@{.}"a"
"a"-@{.}"f" "b"-@{.}"g" "c"-@{.}"h" "d"-@{.}"i" "e"-@{.}"j"
"j"-@{.}"i" "i"-@{.}"h" "h"-@{.}"g" "g"-@{.}"f" "f"-@{.}"j"
"a":@/^/"f" "f":@/^/"a"
"e":@/^/"j" "j":@/_/"i" "i":@/^/"d" "d":@/^/"e"
"g":@/^/"h" _{d} ^{\mathbf{C}} "h":@/_/"c" _{c} "c":@/_/"b" _{b} "b":@/_/"g" _{a}
}
}
\subfloat[]{
\xygraph{
!{<0cm,0cm>;<0cm,1.4cm>:<1.4cm,0cm>::}
!{(0,0);a(0)**{}?(0.6)}*{\bullet}="f"
!{(0,0);a(72)**{}?(0.6)}*{\bullet}="g"
!{(0,0);a(144)**{}?(0.6)}*{\bullet}="h"
!{(0,0);a(216)**{}?(0.6)}*{\bullet}="i"
!{(0,0);a(288)**{}?(0.6)}*{\bullet}="j"
!{(0,0);a(0)**{}?(1.5)}*{s}
!{(0,0);a(0)**{}?(1.3)}*{\bullet}="a"
!{(0,0);a(72)**{}?(1.3)}*{\bullet}="b"
!{(0,0);a(144)**{}?(1.3)}*{\bullet}="c"
!{(0,0);a(216)**{}?(1.3)}*{\bullet}="d"
!{(0,0);a(288)**{}?(1.3)}*{\bullet}="e""a"-@{.}"b" "b"-@{.}"c" "c"-@{.}"d" "d"-@{.}"e" "e"-@{.}"a"
"a"-@{.}"f" "b"-@{.}"g" "c"-@{.}"h" "d"-@{.}"i" "e"-@{.}"j"
"j"-@{.}"i" "i"-@{.}"h" "h"-@{.}"g" "g"-@{.}"f" "f"-@{.}"j"
"a":@/^/"b" "b":@/^/"c"
"c":@/^/"h" "h":@/_/"g" "g":@/_/"f" "f":@/_/"j"
"j":@/^/"e" "e":@/^/"a" 
"d":@/_/"i" _{b} ^{\mathbf{C}} "i":@/_/"d" _{a}
}}
\subfloat[]
{
\xygraph{
!{<0cm,0cm>;<0cm,1.4cm>:<1.4cm,0cm>::}
!{(0,0);a(0)**{}?(0.6)}*{\bullet}="f"
!{(0,0);a(72)**{}?(0.6)}*{\bullet}="g"
!{(0,0);a(144)**{}?(0.6)}*{\bullet}="h"
!{(0,0);a(216)**{}?(0.6)}*{\bullet}="i"
!{(0,0);a(288)**{}?(0.6)}*{\bullet}="j"
!{(0,0);a(0)**{}?(1.5)}*{s}
!{(0,0);a(0)**{}?(1.3)}*{\bullet}="a"
!{(0,0);a(72)**{}?(1.3)}*{\bullet}="b"
!{(0,0);a(144)**{}?(1.3)}*{\bullet}="c"
!{(0,0);a(216)**{}?(1.3)}*{\bullet}="d"
!{(0,0);a(288)**{}?(1.3)}*{\bullet}="e"
"a"-@{.}"b" "b"-@{.}"c" "c"-@{.}"d" "d"-@{.}"e" "e"-@{.}"a"
"a"-@{.}"f" "b"-@{.}"g" "c"-@{.}"h" "d"-@{.}"i" "e"-@{.}"j"
"j"-@{.}"i" "i"-@{.}"h" "h"-@{.}"g" "g"-@{.}"f" "f"-@{.}"j"
"a":@/^/"b" "b":@/^/"c"
"c":@/^/"h" "h":@/_/"g" "g":@/_/"f" "f":@/_/"j"
"j":@/^/"e" "e":@/^/"a" 
"d":@/^/"i" ^{a} "i":@/^/"d" ^{b} _{\mathbf{C}}
}
}

\caption{Two pairs of labeled non-Hamiltonian cycle covers mapped by $M$ in the proof of Lemma~\ref{lem: cc cancel}. (a) and (b) are dual, the arcs along the cycle $\mathbf{C}$ in (a) are reversed in (b) but keep the same labeling.
The fixed--point free duality works also when the cycle consists of only two vertices as shown in the pair (c) and (d).
Note also that this would not be the case if the labels $a$ and $b$ were allowed to be the empty set, since if so they would coincide.}

\label{fig: lcc}
\end{figure}

In this section we show that restrictions on the graph, the computation ring, and the function $f$ can be imposed so that the resulting summation
in a \name{Labeled Cycle Cover Sum} instance is over the Hamiltonian cycle covers only. The contributions of the non-Hamiltonian cycle covers are canceled 
out. First, we say a directed graph is \emph{bidirected} if it for every arc $uv$ has an arc in the opposite direction, $vu$. 
Second, we let the ring $R$ have characteristic two. Third, for an arbitrarily chosen special vertex $s$,
$f:A\times 2^L\setminus \{\emptyset\} \rightarrow R$ is an $s$-\emph{oriented mirror function} if $f(uv,Z)=f(vu,Z)$ for all $Z$ and all $u\neq s, v\neq s$. This definition asymmetry around the vertex $s$ is
a first precaution to avoid that \emph{everything} cancels. We still want the contributions of the Hamiltonian cycle covers to leave a nonzero result. We will address this aspect further in Section~\ref{sec: hc}. 

The following lemma captures how the non-Hamiltonian cycle covers vanish, which also will imply the nonexistence of false positives in our resulting algorithms.
\begin{lem}
\label{lem: cc cancel}
Given a bidirected graph $D=(V,A)$, a finite set $L$, and special vertex $s\in V$,  let $f$ be an $s$-oriented mirror function with a codomain ring of characteristic two. Then
\begin{displaymath}
\Lambda(D,L,f) = \sum_{H\in hc(D)} \sum_{g:L\twoheadrightarrow H} \prod_{a\in H} f(a,g^{-1}(a)).
\end{displaymath}
\end{lem}
\begin{pf}
Confer the definition of \name{Labeled Cycle Cover Sum} (\ref{def: lcc}).  A labeled cycle cover is a tuple $(C,g)$ with $C\in cc(D)$ and $g:L\twoheadrightarrow C$.
We will argue that the labeled non-Hamiltonian cycle covers can be partitioned into dual pairs such that both cycle covers in every pair contribute the same term to the sum. Since we are working in a ring of characteristic two, all these terms cancel. 
To this end we define a mapping $M$ from the labeled non-Hamiltonian cycle covers onto themselves.

Consider a labeled non-Hamiltonian cycle cover $(C,g)$.  We define $M(C,g)=(C',g')$ as follows.
Let $\mathbf{C}$ be the first cycle of $C$ not passing through $s$. Note that there must exist one since the cycle cover consists of at least two cycles and all cycles are vertex disjoint.  Here first
refers to any fixed order of the cycles.
Let $C'=C$ except for the cycle $\mathbf{C}$ which is \emph{reversed} in $C'$, i.e. every arc $uv\in \mathbf{C}$ is replaced by the arc in the opposite direction $vu$ in $C'$. Note that this arc exists since the graph $D$ is assumed to be bidirected. In the special case when $\mathbf{C}$ consists of only two arcs, $C'$ is identical to $C$.
The function $g'^{-1}$ is identical to $g^{-1}$ on $C\setminus \mathbf{C}$, and is defined by $g'^{-1}(uv)=g^{-1}(vu)$ for all arcs $uv\in \mathbf{C}$. In other words, the reversed arcs preserve their original labeling. Note in particular that $(C,g)\neq M(C,g)$ and $(C,g)=M(M(C,g))$. Hence the mapping $M$ uniquely pairs up the labeled non-Hamiltonian cycle covers (cf. Fig.~\ref{fig: lcc}).

Since $f$ is an $s$-oriented mirror function and has $f(uv,Z)=f(vu,Z)$ for all arcs $uv$ not incident to $s$ and all $Z\in 2^L\setminus\{\emptyset\}$, $(C,g)$ and $M(C,g)$ contribute the same product term to the sum in (\ref{def: lcc}) and hence cancel.
\end{pf}

\subsection{Detecting the Hamiltonian Cycles}
\label{sec: hc}
In the previous section we argued that the non-Hamiltonian cycle covers' contributions to the sum in the \name{Labeled Cycle Cover Sum} cancel if certain
requirements are met. For this to be useful we also need that the Hamiltonian cycle covers don't vanish. 
To this end, it will be instructive to think of the elements of $f$ as nonconstant multivariate polynomials in variables associated with the argument arc and label set.
In particular for elements adjacent to the special vertex $s$, $f(su,X)$ and $f(us,X)$ will not share variables for any $su,us\in A$. This will
ensure that the Hamiltonian cycles oriented in opposite directions will contribute different terms to the sum.

In Section~\ref{sec: bip} and \ref{sec: gen} we will define $f$ such that the associated \name{Labeled Cycle Cover Sum} seen as a polynomial in the underlying variables will have at least one unique monomial per (oriented) Hamiltonian cycle. Moreover, there will be no monomials resulting from non-Hamiltonian cycle covers, as a consequence of Lemma~\ref{lem: cc cancel}.

To detect if the polynomial resulting from the \name{Labeled Cycle Cover Sum} is identically zero (=no Hamiltonian cycles) or not (=at least one Hamiltonian cycle),
we will employ the old fingerprint idea often attributed to Freivalds (see \cite{MR95} for a historical account). We will evaluate the polynomial in a randomly chosen point. 
If the fingerprint result is nonzero we know for sure the polynomial couldn't possibly be the zero polynomial. If the result is zero we guess that so is the polynomial.
The Schwartz-Zippel Lemma (see e.g \cite[p. 165]{MR95}) ensures that with great probability we will succeed:

\begin{lem}[Schwartz-Zippel]
\label{lem: SZ}
Let $P(x_1,x_2,...,x_n)$ be a nonzero $n$-variate polynomial of total degree $d$ over a field $F$. Pick $r_1,r_2,...,r_n\in F$ uniformly at random, then
\[
\mbox{Pr}(P(r_1,r_2,...,r_n)=0)\leq \frac{d}{|F|}
\]
\end{lem}

Note that the algorithms' actual computations in this paper will be over a finite field $GF(2^k)$ for some $k$ after replacing the variables for values. In the analysis in Section~\ref{sec: bip} and \ref{sec: gen} though, the function $f$ will be treated as a multivariate polynomial with coefficients from $GF(2)$.

\subsection{Determinants and Inclusion--Exclusion}
Bj\"orklund in \cite{B10} presented a computation technique which suitably tuned can be used to solve \name{Labeled Cycle Cover Sum} relatively quickly.
It relies on the well-known fact that the determinant of an $n\times n$-matrix $\mathbf{A}$ over a ring of characteristic two coincides with the permanent.
\begin{equation}
\label{eq: det}
\mbox{det}(\mathbf{A})=\mbox{per}(\mathbf{A})=\sum_{\sigma:[n]\rightarrow [n]} \prod_{i=1}^n \mathbf{A}_{i,\sigma(i)}
\end{equation}
The summation is over all permutations $\sigma$ of $n$ elements.

Permanents have a natural interpretation as the sum of weighted cycle covers in a directed graph. Formally let $D=(V,A)$ be a directed graph with weights $w:A\rightarrow R$, and define a $|V|\times|V|$ matrix with rows and columns representing the vertices $V$ 
\begin{displaymath}
{\mathbf{A}}_{i,j}=\left\{\begin{array}{ll} w(ij) & : ij\in A \\ 0 & : \mbox{otherwise} \end{array} \right.
\end{displaymath}
then
\begin{equation}
\label{eq: per-cc}
\mbox{per}(\mathbf{A})=\sum_{C\in cc(D)} \prod_{a\in C} w(a).
\end{equation}

We will see that \name{Labeled Cycle Cover Sum} can be evaluated through a sum of an exponential number of determinants.
To this end we define for every $Z\subseteq L$ the matrices

\begin{equation}
\label{def: M}
{\mathbf{M}_f(Z)}_{i,j}=\left\{\begin{array}{ll} f(ij,Z) & : ij\in A,Z\neq \emptyset \\ 0 & : \mbox{otherwise.} \end{array} \right.
\end{equation}

We introduce a polynomial in an indeterminate $r$, with $r$ aimed at controlling the total rank of the subsets used as labels in our labeled cycle covers. 
\begin{equation}
\label{def: p}
p(f,r)=\sum_{Y\subseteq L} \mbox{det} \left( \sum_{Z \subseteq Y} r^{|Z|}\mathbf{M}_f(Z) \right)
\end{equation}
This polynomial can be thought of as an inclusion--exclusion formula in disguise, which actually computes an associated \name{Labeled Cycle Cover Sum} in characteristic two.

\begin{lem}
\label{lem: det}
For a directed graph $D$, a set $L$ of labels, and any $f:A\times 2^L\setminus \{ \emptyset \}\rightarrow GF(2^k)$
\begin{displaymath}
[r^{|L|}]p(f,r) = \Lambda(D,L,f).
\end{displaymath}
\end{lem}
\begin{pf}
Rewriting the expression for $p(f,r)$~(\ref{def: p}) using the equivalence of the determinant and the permanent in rings of characteristic two (\ref{eq: det}), the cycle cover interpretation of the permanent~(\ref{eq: per-cc}), and the matrices $\mathbf{M}_f$~(\ref{def: M}), we get
\[
p(f,r)=\sum_{Y\subseteq L} \sum_{C\in cc(D)} \sum_{q:C\rightarrow 2^Y\setminus \{\emptyset\}}\prod_{a\in C} r^{|q(a)|}f(a,q(a))
\]
Changing the order of summation, we have
\[
p(f,r)\!=\!\!\!\!\!\!\!\sum_{C\in cc(D)}  \sum_{q:C\rightarrow 2^L\setminus \{\emptyset\}}\sum_{\substack{\bigcup_{a\in C} q(a)\subseteq Y\\ Y \subseteq L}} \prod_{a\in C} r^{|q(a)|}f(a,q(a))
\]
For functions $q:C\rightarrow 2^L\setminus \{\emptyset\}$ such that $\bigcup_{a\in C} q(a)\subset L$, i.e. whose union over the elements doesn't cover all of $L$, we note that the innermost
summation is run an \emph{even} number of times with the same term (there are $2^{|L\setminus \bigcup_{a\in C} q(a)|}$ equal terms). Again, since the ring characteristic is two, these cancel. 
We are left with
\[
p(f,r) = \sum_{C\in cc(D)}  \sum_{\substack{q:C\rightarrow 2^L\setminus \{\emptyset\} \\ \bigcup_{a\in C} q(a)=L}} r^{\sum_{a\in C} |q(a)|}\prod_{a\in C} f(a,q(a))
\]
and in particular, the coefficient of $r^{|L|}$
\[
[r^{|L|}]p(f,r)=\sum_{C\in cc(D)}  \sum_{\substack{q:C\rightarrow 2^L\setminus \{\emptyset\} \\ \bigcup_{a\in C} q(a)=L \\ \forall a\neq b:q(a)\cap q(b)=\emptyset}} \prod_{a\in C} f(a,q(a))
\]
since $\bigcup_{a\in C} q(a)=L$ and $\sum_{a\in C} |q(a)|=|L|$ implies $\forall a\neq b:q(a)\cap q(b)=\emptyset$.
Inverting the function $q$ we  arrive at the  \name{Labeled Cycle Cover Sum} definition (\ref{def: lcc}).
\end{pf}

The above lemma is the base identity enabling a relatively efficient algorithm for computing \name{Labeled Cycle Cover Sum}. The runtime is exponential in the number of labels, but polynomial in the size of the input graph.

\begin{lem}
\label{lem: runtime}
The \name{Labeled Cycle Cover Sum} $\Lambda(D,L,f)$ for a function $f$ with codomain $GF(2^k)$  on a directed graph $D$ on $n$ vertices, and with $2^k>|L|n$, can be computed in $O((|L|^2n+|L|n^{1+\omega})2^{|L|}+|L|^2n^2)$ arithmetic operations over $GF(2^k)$, where
$\omega$ is the square matrix multiplication exponent.
\end{lem}
\begin{pf}
We evaluate the \name{Labeled Cycle Cover Sum} via the identity in Lemma~\ref{lem: det}.  We observe that $p(f,r)$ as a polynomial in $r$ has maximum degree $|L|n$. To recover one of its coefficients (the one for $r^{|L|}$), we need to evaluate the polynomial for $|L|n$ choices of $r$ and use interpolation to solve for the sought coefficient. We can for instance use a generator $g$ of the multiplicative group in $GF(2^k)$ and evaluate the polynomial in the points $r=g^0,g^1,g^2,....,g^{|L|n-1}$. The requirement $2^k>|L|n$ ensures the points are distinct, and hence that the interpolation is possible.
For every fixed $r$, our algorithm begins by tabulating $T(Y)=\sum_{Z\subseteq Y} r^{|Z|}M_f(Z)$ for all $Y\subseteq L$ through Yates' fast zeta transform \cite{Y37} (see also \cite{BHKK07} for a recent treatment) in $O(|L|2^{|L|})$ field operations. Next we evaluate $p(f,r)=\sum_{Y\subseteq L} \mbox{det}(T(Y))$ in $O(n^\omega 2^{|L|})$ operations using the determinant algorithm by Bunch and Hopcroft \cite{BH74}, with $\omega=2.376$ the Coppersmith-Winograd square matrix multiplication exponent \cite{CW90}, and store the value obtained. Summing up the number of field operations required over all $|L|n$ values of $r$, the first part of the runtime bound follows.
Once all values are computed, we employ the $O(|L|^2n^2)$ time Lagrange interpolation.
\end{pf}

\section{The Reduction}
\label{sec: red}
We will reduce \name{Hamiltonicity} to \name{Labeled Cycle Cover Sum}. The overall idea is to partition the vertices of the input undirected graph $G$ into two equal halves. We construct a smaller bidirected graph $D$ on one of the halves, and use the other half as labels in a \name{Labeled Cycle Cover Sum} on $D$. 
An especially simple reduction is possible for bipartite graphs, which we describe next, even though the result will also follow directly from Theorem~\ref{thm: bip}.

\subsection{Warm-up: \name{Hamiltonicity} in Bipartite Graphs}
\label{sec: bip}
We are given an undirected bipartite graph $G=(V_1,V_2,E)$ on $n$ vertices. We describe an $O^*(2^{n/2})$ time algorithm detecting if $G$ is Hamiltonian. We know a Hamiltonian cycle if it exists will alternate vertices from $V_1$ and $V_2$ along the cycle. Thus we can safely assume $|V_1|=|V_2|=n/2$ since otherwise the graph is not Hamiltonian.
We will follow the setup outlined in the previous section and imagine a symbolic \name{Labeled Cycle Cover Sum} describing a multivariate polynomial over a ring of characteristic two.
We fix a special vertex $s\in V_1$ and introduce variables $x_{uv}$ and $x_{vu}$ for every edge $uv\in E$. We equate $x_{uv}=x_{vu}$ except when $u=s$ or $v=s$. For every pair of different vertices $u,v\in V_1$,
we define the set $N(u,v)=\{w\colon w\in V_2, uw\in E, wv\in E\}$.
We set $D=(V_1,F)$ with $F$ including arcs in both direction for every pair of different vertices $u,v\in V_1$ such that $N(u,v)\neq \emptyset$. 
For an arc $uv\in F$, and a vertex $w\in N(u,v)$, we set $f(uv,\{w\})=x_{uw}x_{wv}$. In all other points $f$ is set to zero.

\begin{lem}
\label{lem: bip}
With $G, D, V_2,$ and $f$ defined as above, 
\begin{enumerate}
\item[I] $\Lambda(D,V_2,f)=\sum_{H\in hc(G)} \prod_{uv\in H} x_{uv}$
\item[II] $\Lambda(D,V_2,f)$ is the zero polynomial if and only if $hc(G)=\emptyset$.
\end{enumerate}
\end{lem}
\begin{pf}
I. Since $D$ is bidirected and $f$ is easily seen to be an $s$-oriented mirror function, we have from Lemma~\ref{lem: cc cancel} that
\[
\Lambda(D,V_2,f)=\sum_{H\in hc(D)} \sum_{g:V_2\twoheadrightarrow H} \prod_{a\in H} f(a,g^{-1}(a))
\]
Since $f(a,X)$ is nonzero only when $X$ is a single vertex in $V_2$, we can rewrite the identity as
\[
\Lambda(D,V_2,f)=\sum_{H\in hc(D)} \sum_{q:H\rightarrow V_2} \prod_{a\in H} f(a,q(a))
\]
Here the summation is over all functions $q$ which are one-to-one. Replacing $f$ by its definition we get
\[
\Lambda(D,V_2,f)=\sum_{H\in hc(D)} \sum_{\substack{q:H\rightarrow V_2\\ \forall uv\in H:\\ q(uv)\in N(u,v)}} \prod_{wz\in H} x_{wq(wz)}x_{q(wz)z}
\]
Since $q$ is one-to-one, every vertex in $V_2$ is mapped to by precisely one arc $uv\in F$ on a Hamiltonian cycle in $D$, and we have
\[
\Lambda(D,V_2,f)=\sum_{H\in hc(G)} \prod_{uv\in H} x_{uv}
\]
as claimed (remembering that $hc(G)$ contains all oriented Hamiltonian cycles in an undirected graph $G$).

II. Clearly from I, $\Lambda(D,V_2,f)$ is zero if $G$ is non-Hamiltonian. In the other direction, we argue that every undirected Hamiltonian cycle will contribute two different monomials each to the sum. This is again because of the special vertex $s$. Every monomial term in the summation corresponds to an oriented Hamiltonian cycle in $G$ and the variables of the polynomial are uniquely associated with one edge of the graph. Two monomials resulting from two different undirected Hamiltonian cycles will have some variable the other doesn't have. Thus the only chance of two monomials being identical would be the pair of monomials resulting from the same undirected Hamiltonian cycle in opposite orientations. Since the variables tied to the oppositely directed arcs incident to $s$ are different, these are also unique monomials in the sum.
\end{pf}

\subsubsection{Algorithm and Analysis}
The algorithm repeats the following process, called a run, a number of times linear in $n$:

The setup for Lemma~\ref{lem: bip} shows how to transform the input graph $G=(V_1,V_2,E)$ with $|V_1|=|V_2|=n/2$ into a symbolic \name{Labeled Cycle Cover Sum} $\Lambda(D,L,f)$ on a graph $D$ on $n/2$ vertices and $n/2$ labels $L$. We set $k$ large enough, say $2^k>cn$ for some $c>1$.
Next we evaluate $\Lambda(D,L,f)$ in a randomly chosen assignment point $p$ to the variables over the field $GF(2^k)$ with the algorithm from Lemma~\ref{lem: runtime}.
Lemma~\ref{lem: SZ} tells us it will with probability at least $1-1/c$ result in a nonzero answer if and only if $\Lambda(D,L,f)$ was a nonzero polynomial (and $G$ Hamiltonian). 
If any run results in a nonzero answer, we output that $G$ is Hamiltonian, otherwise not.
Since the algorithm uses a linear number of runs in $n$ the probability of false negatives is brought down to $exp(-\Omega(n))$.

The time to compute $\Lambda(D,L,f)$ in $p$ is dominated by the runtime in Lemma~\ref{lem: runtime} and is in $O^*(2^{n/2})$. Summing over all runs, the total time bound only grows by a factor linear in $n$.

\subsection{The General Case}
\label{sec: gen}
In a general Hamiltonian undirected graph $G=(V,E)$, unlike the bipartite case, we don't know a priori which subset of the vertices will be traversed every other vertex along a Hamiltonian cycle in $G$. Hence it is difficult to partition the vertices in two equal parts of which one could serve as labels as in the previous section.
Still, a uniformly randomly chosen partition $V=V_1\cup V_2$ with $|V_1|=|V_2|$ has with large enough probability a 
property that we can exploit: The number of transitions along a fixed Hamiltonian cycle from a vertex in one part to a vertex in the other part is $n/2$ in expectation. Note that in the bipartite case it was $n$. The vertices in $V_1$ in the bipartite case were handled at a polynomial time cost whereas the ones in $V_2$ came at a price of a factor $2$ each in the runtime. In the same vein, the vertices in $V_1$ followed by a vertex in $V_2$ along a fixed Hamiltonian cycle in the general case will be computationally cheap. To see how, we need to distinguish arcs along the fixed Hamiltonian cycle according to the partition.

For a Hamiltonian cycle $H$, we call arcs connecting adjacent vertex pairs $v_i,v_{i+1}$ along $H$ \emph{unlabeled} by $V_2$ if both $v_i$ and $v_{i+1}$ 
belong to $V_1$. The remaining arcs are referred to as \emph{labeled} by $V_2$. We partition the arcs of $H$ in $\mathcal{L}(H)$ as the set 
of labeled, and $\mathcal{U}(H)$ as the set of unlabeled arcs by $V_2$.
We will use that there aren't too many arcs unlabeled by $V_2$.
Define $hc_{V_2}^m(G)$ as the subset of $hc(G)$ of Hamiltonian cycles $H$ which have precisely $m$ arcs unlabeled by $V_2$ along $H$. 

We introduce variables $x_{uv}$ and $x_{vu}$ for every edge $uv\in E$ such that $u\in V_2$ or $v\in V_2$ (or both). We identify $x_{uv}$ with $x_{vu}$ except when $u=s$ or $v=s$.

We consider a complete bidirected graph $D=(V_1,F)$ and use $V_2$ as some of the labels. In addition to $V_2$ we will add a set $L_m$ of size $m$ of extra labels aimed at handling arcs unlabeled by $V_2$. For each edge $uv$ in $G[V_1]$ and every element $d\in L_m$ we also introduce new 
variables $x_{uv,d}$ and $x_{vu,d}$. Again we let $x_{uv,d}$ coincide with $x_{vu,d}$ except when $u=s$ or $v=s$.  

For two vertices $u,v\in V$, and a nonempty subset $X \subseteq V$ we define $\mathcal{P}_{u,v}(X)$ as the family of all simple paths in $G$ from $u$ to $v$ passing through exactly the vertices in $X$ (in addition to $u$ and $v$). 
For $uv\in F$ and $\emptyset\subset X\subseteq V_2$, we set 
\begin{displaymath}
f(uv,X)=\sum_{P\in \mathcal{P}_{u,v}(X)} \prod_{wz\in P} x_{wz}
\end{displaymath}
For every arc $uv\in F$ such that $uv$ is an edge in $G[V_1]$, and every $d\in L_m$, we set
\begin{displaymath}
f(uv,\{d\})=x_{uv,d}
\end{displaymath}
In all other points $f$ is set to zero.

\begin{lem}
\label{lem: gen}
With $G, D, V_2, \mathcal{U}, \mathcal{L}, m, L_m$ and $f$ defined as above, 
\begin{enumerate}
\item[I] $\Lambda(D,V_2\cup L_m,f)=\sum_{H\in hc_{V_2}^m(G)} \biggl(\sum_{\sigma:\mathcal{U}(H)\rightarrow L_m}\prod_{uv\in \mathcal{U}(H)} x_{uv,\sigma(uv)}\biggr)\biggl(\prod_{uv\in \mathcal{L}(H)} x_{uv}\biggr)$

with $\sigma$ one-to-one.
\item[II] $\Lambda(D,V_2\cup L_m,f)$ is the zero polynomial if and only if $hc_{V_2}^m(G)=\emptyset$.
\end{enumerate}
\end{lem}
\begin{pf}
I. Since $D$ is bidirected and $f$ is an $s$-oriented mirror function, we have from Lemma~\ref{lem: cc cancel} that
\[
\Lambda(D,V_2\cup L_m,f)=\sum_{H\in hc(D)} \sum_{g:V_2\cup L_m\twoheadrightarrow H} \prod_{a\in H} f(a,g^{-1}(a))
\]
A Hamiltonian cycle $H\in hc(G)$ is naturally associated with a Hamiltonian cycle in $D$ by simply omitting the vertices along $H$ which belong to $V_2$.
We want to go the other way, to expand the Hamiltonian cycles in $D$ into Hamiltonian cycles of $G$. To do this we observe that the arcs of a Hamiltonian cycle in $D$ in the sum above either are labeled
by an element of $L_m$ or a nonempty subset of $V_2$, since these are the only subsets of the labels for which $f$ is nonzero.  We extend the definition of labeled and unlabeled arcs (which were defined previously for Hamiltonian cycles in $G$ only).
For a Hamiltonian cycle $H\in hc(D)$ labeled by the function $g:V_2\cup L_m\twoheadrightarrow H$ we say
an arc $uv\in H$ is labeled by $V_2$ if $g^{-1}(uv)\subseteq V_2$, and unlabeled by $V_2$ if $g^{-1}(uv)\in L_m$.

Since every arc unlabeled by $V_2$ along a Hamiltonian cycle consumes exactly one of the $m$ labels in $L_m$, and all labels are used, we have that only the Hamiltonian cycles in $D$ with exactly $m$ arcs unlabeled by $V_2$ leave a nonzero contribution. We expand the summation in all possible labeled and unlabeled arcs of the Hamiltonian cycles in $D$.
We note that a cycle $H$ leaves a nonzero result only if the $m$ arcs unlabeled by $V_2$ along the cycle are also edges in $G$. 
\[
\Lambda(D,V_2\cup L_m,f)=
\sum_{H\in hc(D)} \sum_{\substack{H_U\cup H_L=H\\ H_U\cap H_L=\emptyset\\ |H_U|=m\\ \forall a\in H_U:a\in E}} \Lambda_{H_U}(L_m)\Lambda_{H_L}(V_2)
\]
with
\[
\Lambda_{H_U}(L_m)=\left( \sum_{\sigma:H_U\rightarrow L_m} \prod_{a\in H_U} f(a,\sigma(a)) \right)
\]
and
\[
\Lambda_{H_L}(V_2)=\left( \sum_{g:V_2 \twoheadrightarrow H_L} \prod_{a\in H_L} f(a,g^{-1}(a)) \right) 
\]
Here the summation is over all functions $\sigma$ which are one-to-one. Replacing $f$ by its definition we further expand the inner expressions to
\[
\Lambda_{H_U}(L_m)=\left( \sum_{\sigma:H_U\rightarrow L_m} \prod_{uv\in H_U} x_{uv,\sigma(uv)} \right)
\]
and
\[
\Lambda_{H_L}(V_2)=\left( \sum_{g:V_2 \twoheadrightarrow H_L} \prod_{uv\in H_L} \sum_{P\in \mathcal{P}_{u,v}(g^{-1}(uv))} \prod_{wz\in P} x_{wz} \right)
\]
Every vertex in $V_2$ is mapped to by precisely one arc in $F$ on a Hamiltonian cycle $H$ in $D$.
Rewriting the expression as a sum of Hamiltonian cycles in $G$ we have
\begin{displaymath}
\Lambda(D,V_2\cup L_m,f)=  
\sum_{H\in hc_{V_2}^m(G)} \left( \sum_{\sigma:\mathcal{U}(H)\rightarrow L_m}\prod_{uv\in \mathcal{U}(H)} \!x_{uv,\sigma(uv)} \right) \!\! \left( \prod_{uv\in \mathcal{L}(H)} \!x_{uv} \right) 
\end{displaymath}
as claimed.

II. If the graph $G$ has no Hamiltonian cycles, the sum is clearly zero. For the other direction, we see that each Hamiltonian cycle contributes a set of $m!$ different monomials per orientation of the cycle (one for each permutation $\sigma$), in which there are one variable per edge along the cycle. Monomials resulting from different Hamiltonian cycles thus are different since two different Hamiltonian cycles each has an edge the other has not.  Every pair of oppositely oriented Hamiltonian cycles along the same undirected Hamiltonian cycle also traverse $s$ through oppositely directed arcs. Since the variables tied to the oppositely directed arcs incident to $s$ are different, these are also unique monomials in the sum.
\end{pf}

\subsubsection{Algorithm}
We are given an $n$ vertex undirected graph $G=(V,E)$ as input, assuming $n$ even for simplicity.
The algorithm repeats the following process, called a run, several times $r$ to be specified later in Section~\ref{sec: gen an}:

In each run, a partition $V_1\cup V_2=V$ is picked uniformly at random, with $|V_1|=|V_2|=n/2$. Next the algorithm loops over $m$, the number of edges unlabeled by $V_2$ along a Hamiltonian cycle, from $0$ through $m_{max}$, where $m_{max}$ is specified in Section~\ref{sec: gen an}. For each value of $m$,
the setup for Lemma~\ref{lem: gen} describes how to transform the input graph $G$ given $m,V_1,V_2$ into a symbolic \name{Labeled Cycle Cover Sum} $\Lambda(D,V_2\cup L_m,f)$ on a bidirected graph $D$ on $n/2$ vertices and $n/2+m$ labels $V_2\cup L_m$. We operate over a field $GF(2^k)$ with $k$ again set  large enough, say $2^k>cn$ for some $c>1$.  

Next in each run, a point $p$ assigning values from the field $GF(2^k)$ to the variables in $f$ is chosen uniformly at random.
The function $f$ in $p$ is tabulated for all subsets of $V_2$ (in other points the function is easy to compute). This can be achieved by running a variant of the Bellman-Held-Karp recursion.
Formally, let $\hat{f}:(V\times V)\times 2^{V_2}\rightarrow \mbox{GF}(2^k)$ be defined for $u\neq v$ and $\emptyset \subset X \subseteq V_2$ by 
\begin{displaymath}
\hat{f}(uv,X)=\sum_{P\in \mathcal{P}_{u,v}(X)} \prod_{wz\in P} x_{wz}
\end{displaymath}
then $f(uv,X)=\hat{f}(uv,X)$ for $uv\in F$ and $\emptyset \subset X\subseteq V_2$. 

For $|X|>1$ the recursion
\begin{displaymath}
\hat{f}(uv,X) = \sum_{w\in X,uw\in E} x_{uw}\hat{f}(wv,X\setminus \{w\})
\end{displaymath}
can be used to tabulate $f(.,X)$ for all $\emptyset\subset X\subseteq V_2$. 

Finally, in each run, we evaluate $\Lambda(D,V_2\cup L_m,f)$ in $p$ through Lemma~\ref{lem: runtime}. If the result in any of the runs is nonzero we conclude and output that $G$ is Hamiltonian, otherwise not.

\subsubsection{Analysis}
\label{sec: gen an}

Since the tabulation for $f$ is in $O^*(2^{0.5n})$ time, the runtime of each run is dominated by the runtime in Lemma~\ref{lem: runtime}. The worst case occurs for $m=m_{max}$ in which case we get a $O^*(2^{0.5n+m_{max}})$ time bound. The total runtime is in $O^*(r2^{0.5n+m_{max}})$, 
and the probability of false negatives is at most $Pr(\sum_{m=0}^{m_{max}}|hc_{V_2}^{m}(G)|=0)^r$. 

Lemma~\ref{lem: SZ} tells us that every run will with probability at least $1-1/c$ result in a nonzero answer if and only if $\Lambda(D,V_2\cup L_m,f)$ was a nonzero polynomial (and $G$ has a Hamiltonian cycle with $m$ edges unlabeled by $V_2$).
A straightforward application of Markov's inequality bounding the probability that a fixed Hamiltonian cycle gets more than $n/4$ edges unlabeled by $V_2$ shows that
$\mbox{Pr}(\sum_{m=0}^{n/4}|hc_{V_2}^m(G)|=0)\leq n/(4+n)$. From this we can deduce that setting $m_{max}=n/4$ and $r=\Omega(n^2)$, will give a total runtime bound of $O^*(2^{\frac{3}{4}n})\approx O^*(1.682^n)$ and exponentially small probability of failure in $n$.

A slightly better runtime bound is obtained by trading the probability of success in a single run for more runs (suggested to the author by Ryan Williams and Petteri Kaski, independently of each other).
To see just how much better we need a stronger bound on the probability.

\begin{lem}
\label{lem: unl}
Let $G=(V,E)$ be a Hamiltonian undirected graph, and $V_1\bigcup V_2=V$ with $|V_1|=|V_2|=n/2$. Then
\[
\mbox{Pr}(|hc_{V_2}^m(G)|>0) \geq \frac{ \binom{n/2-1}{m-1}^2}{\binom{n}{n/2}} \in \Theta\!\! \left( \frac{m \left(\frac{1}{2}-\frac{m}{n}\right)^{2m-n-1} }{\left(\frac{m}{n}\right)^{2m}4^n\sqrt{2\pi} n^{1.5}} \right)
\]
\end{lem}
\begin{pf}
We will obtain the first bound by counting the probability that one fixed Hamiltonian cycle $H=(v_0,v_1,\cdots, v_{n-1})$ has exactly $m$ arcs unlabeled by $V_2$, and moreover has $v_0\in V_1$ and $v_{n-1}\in V_2$.
For such a $H$ there are exactly $n/2-m$ indices $i$ such that $x_{i}\in V_1$ and $x_{i+1}\in V_2$, and just as many indices $i$ where $x_{i}\in V_2$ and $x_{i+1}\in V_1$.
The ordered list of these transition indices $i_1,i_2,\cdots,i_{n-2m}$ uniquely describes the partition, and in the other direction any such list with $0\leq i_1$, $\forall j:i_j<i_{j+1}$, and $i_{n-2m}= n-1$ corresponds to a unique partition. Set $i_0=-1$ and define the positive integers $d_j=i_j-i_{j-1}$ for all $0<j\leq n-2m$.
Note that $d_1,d_3,\cdots, d_{n-2m-1}$ describes a partition of the vertices in $V_1$ in $n/2-m$ groups. Analogously, $d_2,d_4,\cdots, d_{n-2m}$ describes a partition of the vertices in $V_2$ in $n/2-m$ groups. The number of ways to write a positive integer $p$ as a sum of $k$ positive integers is $\binom{p-1}{k-1}$. Multiplying the number of ways to partition the vertices in $V_1$ with the number of ways for $V_2$, and dividing with the total number of balanced partitions $\binom{n}{n/2}$, the first result follows.

The second bound is derived by replacing the binomial coefficients with their factorial definition and using Stirling's approximation for $n!\in \Theta((n/e)^n\sqrt{2\pi n})$.
\end{pf}

To bring the probability of false negatives down to $exp(-\Omega(n))$, we need the number of runs $r=n^{O(1)}Pr^{-1}(|hc_{V_2}^{m_{max}}(G)|>0)$. Solving for the local minimum of the total runtime, using Lemma~\ref{lem: unl} to bound the probability, we get $m_{max}=0.205$ and $r=n^{O(1)}2^{0.024n}$ runs. Altogether, a runtime bound of $O^*(1.657^n)$.

For Theorem~\ref{thm: bip}, we set $V_2$ equal to the independent set of size $i$ given as input instead of the random partition in the first step 
of the algorithm above. We note that a Hamiltonian cycle must have at least $2i$ arcs labeled by $V_2$. 
This is because every vertex of the independent set must be incident to two arcs along the cycle, and no arc is connected to more than one of them since they are disconnected by definition. Hence, if $i>n/2$  the graph is surely non-Hamiltonian, and if $i\leq n/2$ we only need the loop over $m$ in the above algorithm to count to $m_{max}=n-2i$, and let $r$ be linear in $n$. This gives us the $O^*(2^{n-i})$ time bound.

\section{Polynomial Space}
\label{sec: pspace}
The algorithm in Lemma~\ref{lem: runtime} invokes Yates' fast zeta transform which uses almost as much space as time. 
The tabulation of the function $f$ also uses space exponential in $n$.
Here we describe an alternative way of solving the problem with an algorithm using only polynomial space. This also enables the summation task in the \name{Labeled Cycle Cover Sum} to be divided on several processors in parallel. We adopt the notation from Section~\ref{sec: gen}.
There we described $f$ as a function of simple paths in $G[V_2]$.
The idea here is to embed the inclusion--exclusion counting over walks to sieve for the simple paths. We let $\mathcal{W}_{u,v}(X,l)$ be the set of walks of length $l$ in $G$ starting in $u$ and ending in $v$ but in-between visiting only vertices from $X$. For a walk $W\in \mathcal{W}_{u,v}(X,l)$ we define its support on $X$, denoted $S_X(W)$, as the set of vertices in $X$ traversed by the walk. With $W=(u,w_1,w_2,\dotsc, w_{l-1}, v)$ we have $S_X(W)=\bigcup_{i=1}^{l-1} w_i$.

Note that $\mathcal{P}_{u,v}(X)\subseteq \mathcal{W}_{u,v}(X,|X|+1)$. 
We will derive an analogue $g$ of the function $f$. We replace simple paths for walks. The point being that the inner summation in $p(f,r)$ in (\ref{def: p}) will take the form of counting over walks, which we know how to do fast using only polynomial space.

For an arc $uv\in F$, and a subset $\emptyset\subset X \subseteq V_2$ we set 
\begin{displaymath}
g(uv,X,r)=\sum_{k=|X|}^n r^k \sum_{\substack{W\in \mathcal{W}_{u,v}(X,k+1)\\ S_X(W)=X}} \prod_{wz\in W} x_{wz}
\end{displaymath}
For every arc $uv\in F$ such that $uv$ is an edge in $G[V_1]$, and every $d\in L_m$, we set
\begin{displaymath}
g(uv,\{d\},r)=rx_{uv,d}
\end{displaymath}
In all other points $g$ is set to zero. 

Remembering the definition of $\mathbf{M}$ from (\ref{def: M}), we define an analogue of $p(f,r)$ from (\ref{def: p}).
\begin{equation}
\label{def: q}
q(g,r)=\sum_{Y\subseteq L} \mbox{det}\left( \sum_{Z \subseteq Y} \mathbf{M}_{g(.,.,r)}(Z) \right)
\end{equation}

\begin{lem}
\label{lem: det-poly}
For an undirected graph $G=(V,E)$, a vertex partition $V_1\cup V_2=V$, and with $f$ defined in Section~\ref{sec: gen} and $g$ defined as above, it holds that
\begin{displaymath}
[r^L]q(g,r) = [r^L]p(f,r)
\end{displaymath}
\end{lem}

\begin{pf}
Rewriting the expression for $q(g,r)$ (\ref{def: q}) using the equivalence of the determinant and the permanent in rings of characteristic two (\ref{eq: det}), the cycle cover interpretation of the permanent (\ref{eq: per-cc}), and the matrices $\mathbf{M}_{g(.,.,r)}$ (\ref{def: M}), we get
\[
q(g,r) = \sum_{Y\subseteq L} \sum_{C\in cc(G)} \sum_{h:C\rightarrow 2^Y\setminus \{\emptyset\}}\prod_{a\in C} g(a,h(a),r)
\]
Changing the order of summation, we have
\[
q(g,r)\!=\!\!\!\!\! \sum_{C\in cc(G)}  \sum_{h:C\rightarrow 2^L\setminus \{\emptyset\}}\sum_{{\bigcup_{a\in C} h(a)\subseteq Y\subseteq L}} \prod_{a\in C} g(a,h(a),r)
\]
For functions $h:C\rightarrow 2^L\setminus \{\emptyset\}$ such that $\bigcup_{a\in C} h(a) \subset L$, i.e. whose union over the elements doesn't cover all of $L$, we note that the innermost
summation is run an \emph{even} number of times with the same term (there are $2^{|L\setminus \bigcup_{a\in C} h(a)|}$ equal terms). Again, since the characteristic is two, these cancel.
\[
q(g,r) = \sum_{C\in cc(G)}  \sum_{\substack{h:C\rightarrow 2^L\setminus \{\emptyset\} \\ \bigcup_{a\in C} h(a)=L}} \prod_{a\in C} g(a,h(a),r)
\]
Restricted to the monomial for $r^{|L|}$, the only monomial (in $r$) in any $g(a,X,r)$ which contributes to the final sum is the one for $r^{|X|}$, since there are no nonzero monomials of smaller degree in $g(a,X,r)$, and the total degree of the monomial product should be $L$. We have
\[
[r^{|L|}]q(g,r)\!=\!\!\!\!\! \sum_{C\in cc(G)}  \sum_{\substack{h:C\rightarrow 2^L\setminus \{\emptyset\} \\ \bigcup_{a\in C} h(a)=L \\ \sum_{a\in C} |h(a)|=|L|}} \prod_{a\in C}  [r^{|h(a)|} ] g(a,h(a),r)
\]
Since $[r^{|X|}]g(a,X,r)=f(a,X)$ and $\bigcup_{a\in C} h(a)=L$ together with $\sum_{a\in C} |h(a)|=|L|$ implies $\forall a\neq b:h(a)\cap h(b)=\emptyset$, we get
\[
[r^{|L|}]q(g,r)=\sum_{C\in cc(G)}  \sum_{\substack{h:C\rightarrow 2^L\setminus \{\emptyset\} \\ \bigcup_{a\in C} h(a)=L \\ \forall a\neq b:h(a)\cap h(b)=\emptyset}} \prod_{a\in C} f(a,h(a))
\]
which is the same expression as for $[r^{|L|}]p(f,r)$ at the end of the proof of Lemma~\ref{lem: det}.
\end{pf}

The above lemma offers an alternative route to compute $[r^{|L]}]p(f,r)$ by evaluating the coefficient of $r^{|L|}$ in $q(g,r)$ instead. To achieve this, we again compute $q(g,r)$ for $|L|n$ values on $r$ and use Lagrange interpolation to retrieve the coefficient. To compute $q(g,r)$ for a fixed $r$, we don't tabulate the values of $g$ as we did for $f$ in the evaluation of $p(f,r)$. Instead we note that the inner sum in (\ref{def: q}), $\sum_{Y\subseteq X} \mathbf{M}_{g(.,.,r)}(Y)$, can be evaluated in time and space polynomial in $|X|$. To see how, let $X_1 = X \cap V_2$ and $X_2 = X \cap L_m$ and note that 
\begin{equation}
\label{eq: dec}
\sum_{Y\subseteq X} \mathbf{M}_{g(.,.,r)}(Y)=\!\!\sum_{Y\subseteq X_1} \mathbf{M}_{g(.,.,r)}(Y)+\sum_{Y\subseteq X_2} \mathbf{M}_{g(.,.,r)}(Y)
\end{equation}
This decomposition is valid since $g(a,Y,r)$ takes by definition the value $0$ for all $Y$ such that both $Y\cap X_1\neq \emptyset$ and $Y\cap X_2\neq \emptyset$.
The second sum in the rhs of (\ref{eq: dec}) is easily evaluated since $g(a,Y,r)$ also by definition takes the value zero for all $Y$ such that $|Y\cap X_2|>1$.
In the first sum of the rhs of (\ref{eq: dec}) the coefficient of $r^l$ in row $u$ and column $v$ in the resulting matrix equals zero if $u=v$ and otherwise
evaluates to 
\begin{equation}
\label{eq: uv}
\sum_{Y\subseteq X_1}\sum_{\substack{W\in \mathcal{W}_{u,v}(Y,l+1)\\ S_X(W)=Y}} \prod_{wz\in W} x_{wz}\! =\!\!\! \!\sum_{W\in \mathcal{W}_{u,v}(X_1,l+1)} \prod_{wz\in W} x_{wz}
\end{equation}
i.e. the number of weighted walks from $u$ to $v$ passing through $l$ vertices (possibly with repetition) in $X_1$.
These can be evaluated efficiently. Simple let an $|X_1|\times |X_1|$ matrix $\mathbf{A}$ with rows and columns representing vertices of $X_1$ be defined by 
\begin{displaymath}
\mathbf{A}_{u,v}=\left\{\begin{array}{ll} x_{uv} & : uv\in E \\ 0 & : \mbox{otherwise} \end{array} \right.
\end{displaymath}
and an $|V_1|\times |X_1|$ matrix $\mathbf{B}$ with rows representing vertices of $V_1$ and columns vertices of $X_1$ by
\begin{displaymath}
\mathbf{B}_{u,v}=\left\{\begin{array}{ll} x_{uv} & : uv\in E \\ 0 & : \mbox{otherwise} \end{array} \right.
\end{displaymath}
Then, the element at row $u$ and column $v$ of the matrix product $\mathbf{B}\mathbf{A}^{l}\mathbf{B}^{\mbox{\tiny{{T}}}}$ by matrix multiplication definition equals
the rhs of (\ref{eq: uv}). This product can trivially be computed in time and space polynomial in $n$.

\section{TSP with Bounded Integer Weights}
\label{sec: TSP}
In the TSP edges have weights $\ell :E\rightarrow \mathbb{Z^+}$ and we seek the Hamiltonian cycle with the smallest total weight. Since our approach is algebraic, in particular operating on
a sum--product ring, it is not evident how to handle a minimum query efficiently. For small weights though, there is an embedding solution. 
We note that the monomials in the polynomials we evaluate is a product over all edges along a Hamiltonian cycle. We introduce yet an 
auxiliary indeterminate $y$, meant to sort the Hamiltonian cycles after total weight. For every actual edge $uv$ we represent, we also multiply it with $y^{\ell(uv)}$. Formally, we extend the definition of $f$ from Section~\ref{sec: gen}.
For $uv\in F$ and $\emptyset\subset X\subseteq V_2$, we set 
\begin{displaymath}
f_y(uv,X)=\sum_{P\in \mathcal{P}_{u,v}(X)} \prod_{wz\in P} y^{\ell(wz)}x_{wz}
\end{displaymath}
For every arc $uv\in F$ such that $uv$ is an edge in $G[V_1]$, and every $d\in L_m$, we set
\begin{displaymath}
f_y(uv,\{d\})=y^{\ell(uv)}x_{uv,d}
\end{displaymath}

Now $\sum_{i=0}^{m_{max}}\Lambda(D,V_2\cup L_i,f_y)$ seen as a polynomial in $y$, has a nonzero coefficient for the monomial $y^l$ only if there is a Hamiltonian cycle of total weight $l$. 
This idea of embedding a min-sum semi-ring on a sum-product ring is quite old. It was used by Kohn et al.~\cite{KGK77} for TSP, and earlier by Yuval~\cite{Y76} for all-pairs shortest paths.

We can use the Fast Fourier Transform on $GF(2^k)$ to retrieve the smallest $l$ for which the coefficient of $y^l$ is nonzero. 
First of all, we need to have $2^k$ larger than the maximum degree of a monomial in the indeterminate $y$ to avoid having monomial coefficients wrap around $y^{2^k-1}$. We note that the largest degree 
equals the weight of the heaviest Hamiltonian cycle. This weight at least is less than $w$, the sum of all weights, and we set $2w\geq2^k>w$.
For a generator $g$ of the multiplicative group in $GF(2^k)$, we compute and tabulate $T(l)=\sum_{i=0}^{m_{max}}\Lambda(D,V_2\cup L_i,f_y)$ evaluated at $y=g^l$ for every $l=0,1,\cdots,2^k-2$.
The runtime of this step is in $O^*(w2^{0.5n+m_{max}})$. Next we compute the inverse Fourier transform of $T$:
\begin{displaymath}
t(j)=\sum_{l=0}^{2^k-2} g^{-jl}T(l) 
\end{displaymath}
The value $t(j)$ equals the coefficient of $y^j$ in $\sum_{i=0}^{m_{max}}\Lambda(D,V_2\cup L_i,f_y)$, since for $k>1$
\begin{displaymath}
\sum_{l=0}^{2^k-2} g^{-jl}g^{il}=\left\{\begin{array}{ll} 1 & : i=j \\ 0 & : i\neq j \end{array} \right.
\end{displaymath}
By the Fast Fourier Transform, $t(j)$ is computed for all $j$ in $O(w\log w)$ time. Finally, we search linearly in $t(j)$
to find the smallest $j$ such that the coefficient of $y^j$ is nonzero. 

To increase the probability of the lightest cycle to show in the sum (remembering that we only detect Hamiltonian cycles with at most $m_{max}$ arcs unlabeled by $V_2$ this way),
we rerun the algorithm $r$ times with different partitions $V_1\cup V_2=V$ as before. We output the smallest $j$ found for which the coefficient of $y^j$ was nonzero in any run.
This proves Theorem~\ref{thm: TSP}.

\section*{Acknowledgment}
The author is grateful to an anonymous referee, Thore Husfeldt, Petteri Kaski, Mikko Koivisto, and Ryan Williams for extensive commenting on an earlier draft of the paper.
This work was supported in part by the Swedish Research Council project ``Exact Algorithms''.

\end{document}